# Electroluminescence from indirect band gap semiconductor $ReS_2$

Ignacio Gutiérrez-Lezama,* Bojja Aditya Reddy, Nicolas Ubrig and Alberto F. Morpurgo*

DQMP and GAP, Université de Genève, 24 quai Ernest Ansermet, CH-1211 Geneva, Switzerland

**Abstract**

It has been recently claimed that bulk crystals of transition metal dichalcogenide (TMD) $ReS_2$ are direct band gap semiconductors, which would make this material an ideal candidate, among all TMDs, for the realization of efficient opto-electronic devices. The situation is however unclear, because even more recently an indirect transition in the PL spectra of this material has been detected, whose energy is smaller than the supposed direct gap. To address this issue we exploit the properties of ionic liquid gated field-effect transistors (FETs) to investigate the gap structure of bulk $ReS_2$. Using these devices, whose high quality is demonstrated by a record high electron FET mobility of 1,100 $cm^2/Vs$ at 4K, we can induce hole transport at the surface of the material and determine quantitatively the smallest band gap present in the material, irrespective of its direct or indirect nature. The value of the band gap is found to be 1.41 eV, smaller than the 1.5 eV direct optical transition but in good agreement with the energy of the indirect optical transition, providing an independent confirmation that bulk $ReS_2$ is an indirect band gap semiconductor. Nevertheless, contrary to the case of more commonly studied semiconducting TMDs (e.g., $MoS_2$, $WS_2$, etc.) in their bulk form, we also find that $ReS_2$ FETs fabricated on bulk crystals do exhibit electroluminescence when driven in the ambipolar injection regime, likely because the difference between direct and indirect gap is only 100 meV. We conclude that $ReS_2$ does deserve more in-depth investigations in relation to possible opto-electronic applications.

**Introduction**

Layered semiconducting transition metal dichlcogenides (TMDs), such as (2H) $MoS_2$, $WS_2$, $MoSe_2$, $WSe_2$, and $MoTe_2$, are of interest for the realization of opto-electronic devices[1–7], because they support ambipolar transport with balanced electron and hole mobility[5,8–10], and because –upon changing the compound and varying the crystal thickness – they exhibit radiative transitions spanning a broad range of energy[11–16]. One aspect that is crucial in the context of optoelectronic applications is the nature of the band gap, which needs to be direct to ensure that coupling with light is sufficiently strong. For all materials listed above, the band gap is direct only when their thickness is reduced to the monolayer level[11–13] (bilayer, in the case of $2H-MoTe_2$[16]), which –albeit of great interest in the study of a number of physical phenomena– is not optimal for practical applications. Indeed, it would be highly desirable to have the possibility to access all positive aspects of semiconducting TMDs in a material having direct band gap irrespective of its thickness, i.e. also in its bulk form. In this context, the recent report[17] that bulk $ReS_2$ in its distorted 1T phase is a direct band gap semiconductor has attracted considerable attention. However, the situation is now unclear because an even more recent study[18] combining photoluminescence (PL) and reflectivity contrast measurements reported the observation of an optical transition in bulk $ReS_2$ having indirect character and lower energy than the one associated to the direct gap. As the matter is particularly relevant for possible technological developments, it is important to clarify the situation.

Here we use ionic liquid (IL) field-effect transistors (FETs) to determine the gap of bulk $ReS_2$ quantitatively. Owing to their extremely large electrostatic capacitance, IL gated FETs allow the quantitative determination of the band gap of a semiconductor in a conceptually straightforward manner if ambipolar transport –i.e., both electron and hole conduction- can be induced upon sweeping the gate voltage $V_G$[19]. In this case, the gap value is directly obtained from the difference in threshold voltage for electron and hole conduction, which correspond to the energy needed to add an electron to the bottom of the conduction band and to create a hole at the top of the valence band. As we have shown in the past on a variety of materials (bulk[19], mono and bilayer $WS_2$[5], bulk $MoTe_2$[10], and monolayer $MoS_2$[20]), the technique is reliable and allows the value of the gap to be determined precisely (the precision can be better than 5% if the device electrical characteristics are sufficiently good). We emphasize that the technique gives the actual gap irrespective of its direct or indirect nature, and not the exciton recombination energy (the difference between the two is the exciton binding energy, which is normally small in bulk materials owing to electrostatic screening, but it can become a large fraction of 1 eV when the material thickness is reduced to that of an individual monolayer)[21]. The use of ambipolar IL FETs to determine the gap offers additional advantages, since the same devices that are realized to implement this technique can be employed to search for electroluminescence (EL)[5,6,20]. This is particularly useful in the present case, since the observation of EL gives indications as to the potential of the material for opto-electronic applications.

We find that IL gated FETs realized on bulk $ReS_2$ do exhibit ambipolar transport, and from the devices transfer characteristics we determine the band gap of $ReS_2$ crystals to be 1.41± 0.05 eV. This value is smaller than the 1.5 eV direct transition that has been claimed to be the direct band gap[17], and is in very good agreement with the smallest optical transition observed in the PL spectra of thick (>150 nm) $ReS_2$ crystals, corresponding to the indirect transition recently reported in ref. [18]. Bulk $ReS_2$ is therefore an indirect band gap semiconductor. Using the same devices, we further show that –despite the indirect nature of the band gap– light is emitted when the devices are operated in the ambipolar injection regime, i.e. when electrons and holes are injected simultaneously at opposite contacts. A comparison of the spectra in $ReS_2$ crystals reveals that the radiative process observed in EL corresponds to the 1.5 eV direct transition detected in PL. This finding is remarkable because in ambipolar ionic-liquid gated FETs realized using bulk crystals of the more common semiconducting TMD compounds ($MoS_2$, $WS_2$, etc.) no light emission is observed, since for these compounds EL is only present when their thickness is reduced to a few atomic layers. EL is visible in ReS$_2$ because the difference in energy between the direct and indirect gap is only approximately 100 meV, i.e., much smaller than in bulk crystals of other more common TMDs (where it is 0.6 - 0.8 eV depending on the compound). For this reason we conclude that bulk ReS$_2$ may have technological relevance already in its bulk form and its properties should be investigated in more depth, especially focusing in their evolution as a function of thickness.

**Experimental**

The devices discussed here were realized using ReS$_2$ crystals whose thickness typically ranged between 15 and 25 nm, behaving in all regards as bulk, obtained by mechanically exfoliating macroscopically thick crystals purchased from 2D-semiconductors Inc. The exfoliated layers were transferred onto highly p-doped Si substrates covered by SiO$_2$ (285 nm thick). The doped substrate could be used as a back gate, which proved useful to characterize the material properties and the effect of the IL gate (see next section). Au contacts defining a Hall-bar geometry were realized using conventional nano-fabrication techniques (e-beam lithography, e-beam evaporation, and lift-off; see figure 1(a)). After deposition, the contacts were thermal annealed for 3 hours at 480 K in a stream of H and Ar gas. When possible, the devices were further annealed *in-situ*, in the chamber of the variable temperature insert used to perform electrical measurements, in high vacuum at 350 K. We found that this *in-situ* annealing process repeatedly led to much lower OFF-currents in the FETs electrical characteristics, as compared to devices that had only been annealed *ex-situ*. Along with the contacts, a large-area Au pad (the gate electrode in the presence of the ionic liquid) and a reference electrode were deposited onto the substrate (see figure 1(b)) for a scheme of the devices).

All the transport measurements were performed either in a variable temperature insert operating in the temperature range between 350 and 2 K, equipped with a 7T magnet, or in the vacuum chamber of a continuous-flow cryostat with optical access, in which the devices can be cooled down from 300 K to 4K by the continuous flow of liquid helium through a cold finger. The electrical measurements were performed along the high-electron mobility direction (b-axis) of the ReS$_2$ crystals (known to be the direction parallel to the cleaved edges) in a voltage-bias configuration using home-made low-noise current sources and voltage amplifiers. Two ionic liquids were used, DEME-TFSI and F14-FAP, with no observed difference in the obtained results. The measurements performed with the SiO$_2$/Si gate were done in the absence of IL; the SiO$_2$/Si gate was left floating during the IL-gated measurements. PL and EL experiments were carried out in a home-made confocal micro-PL setup with the devices mounted in the above mentioned continuous-flow cryostat with optical access. The PL measurements were performed in a back-scattering geometry (i.e., collecting the emitted light with the same objective used to couple the laser beam onto the device) at an excitation wavelength of 488 nm. In PL and EL the light collected from the sample was sent to a Czerny-Turner monochromator (Andor Shamrock5001) and detected with a thermoelectric cooled Si CCD array (Andor Newton CCD).

**Results and discussion**

The use of an IL-gated FET to extract the band gap of a semiconducting material relies on the ability to observe ambipolar transport characteristics of sufficient quality, which generally depends on a number of aspects specific to the material considered (see, for instance, ref. [10]). In the particular case of ReS$_2$ devices it is not a priori obvious that the method may work because, owing to the very narrow band-width of the material[17,22], the use of IL gate may cause a drastic suppression of the carrier mobility, as it has been recently observed at high applied gate voltages in polymer electrolyte gated ReS$_2$ FETs[23]. Indeed, as for other narrow band semiconductors (0.3 eV), such as organic crystals[24], the disorder induced by the random Coulomb potential generated by ions in the liquid next to the semiconductor surface can strongly increase the level of disorder and drastically suppress transport in ReS$_2$ crystals. It is therefore important to investigate if the presence of the IL has a detrimental effect on the I-V characteristics of our ReS$_2$ FETs in the region just above threshold, which is the one relevant for the determination of the band gap. To this end we have used the SiO$_2$/Si (back) gate present in our devices to assess their quality prior to the IL gate measurements, and to compare the performance of the devices operated with the two different gates. Since ReS$_2$ is generally found to be *n*-doped[22,23,25,26] and hole transport cannot be induced using a Si/SiO$_2$ gate, for this comparison we focus on electron transport.

Figure 1(c) shows the transfer curves (drain-source current $I_{DS}$ versus gate voltage $V_G$) measured at high temperature (250K) as a function of the gate voltage applied to either the IL (top) gate or to the SiO$_2$/Si (back) gate, for electron densities ranging up to approximately $10^{13}$ cm$^{-2}$ (as determined independently from Hall effect measurements). Figure 1(d) shows the same comparison for the conductivity extracted from four terminal measurements, to eliminate the effect of the contact resistance and to account for the device geometry. Except for the very different efficiency of the gate voltage (due to the different gate capacitances), the overall behavior is virtually identical. At a more quantitative level, figure 1(e) compares the carrier density dependent electron mobility extracted for the two cases, with electrons accumulated either with the IL or with the back gate (the transistor mobility obtained using the known value of the back-gate capacitance –rather than the electron density determined by the Hall effect– is also plotted). All the mobility values coincide, within better than a factor of two, showing that in this range of carrier densities disorder induced by the IL is not posing problems to the measurements.

As an additional characterization of the ReS$_2$ crystals, we have probed the transport properties as a function of temperature, down to 4.2 K. As the ionic liquids used here (see methods section) freeze at relatively high temperature (just under 200 K), for these measurements we have relied on the Si/SiO$_2$ back gate. As it can be seen from the gate-dependent four terminal conductivity $\sigma_{4T}$ (figure 1(f)), at sufficiently high gate voltage ($V_G$ = 80 V) the current is found to increase steadily upon cooling, indicating the occurrence of metallic transport throughout the entire temperature range investigated. This conclusion is reiterated in the inset of figure 1(f), which shows the temperature dependence of $\sigma_{4T}$ measured for three different values of $V_G$. Unlike previous reports, which always found a downturn of the conductivity at low temperature[25–27], in our devices at high $V_G$ metallic behavior can be seen throughout the whole investigated temperature range. From the measurements we can also extract the temperature dependent FET electron mobility $\mu = 1/C \, (d\sigma_{4T}/dV_G)$ (where $C$ is the capacitance per unit area of the SiO$_2$ gate insulator) and find that $\mu$ reaches a value as high as 1,100 cm$^2$/Vs at 4K for $V_G = 80$ V, higher than the electron mobility values reported earlier for ReS$_2$[25,26]. We conclude that, as far as it can be inferred from these tests, both the material and our devices are of high quality, at least when electron transport properties are concerned.

As the occurrence of ambipolar conduction is a necessary condition for the measurement of the band gap and light emission, we now turn our attention to the investigation of hole transport by using the IL gate. Figure 2(a) shows the transfer curve of the same IL-gated FET whose characteristics were discussed above, as a function of reference potential $V_{REF}$ measured with a reference electrode immersed in the ionic liquid (see the device scheme in figure 1(a) and the experimental section for more details) and of $V_G$ (inset of figure 2(a)). Hole transport is clearly observed when a sufficiently large and negative gate voltage $V_G$ is applied. Interestingly, when the data are plotted as a function of reference potential $V_{REF}$ –which corresponds to the potential of the IL relative to ground (or, equivalently when the FET is ON, to semiconductor surface)– virtually no hysteresis is seen in the transfer curve, even though a significantly larger hysteresis is seen in the transfer curve plotted as a function of $V_G$. Consistently with this observation, we find that hysteresis is present in the $V_{REF}$-vs-$V_G$ curve (Figure 2(b)). We conclude that what is hysteretic is the relation between the applied gate voltage and the potential inside the ionic liquid, and not the behavior of our FET devices. In other words, these observations establish that the hysteresis originates from processes occurring at the interface between the gate and the IL, and not at the interface between the IL and the surface of the semiconductor.

As compared to electrons, holes in ReS$_2$ exhibit 10 – 50 times smaller conductivity (depending on the specific device). Under hole accumulation, we could not determine the carrier density by means of Hall effect measurements, which were impeded by bias stress effects –i.e. the drift of the threshold voltage upon continued gate bias. At very large negative gate voltage these effects become pronounced and the resulting drift makes the devices insufficiently stable to measure their magneto-resistance. Nevertheless, if we assume that the capacitance per unit area of the devices at large negative gate voltage has the same order of magnitude as the capacitance estimated under electron accumulation, we conclude that the smaller hole conductivity is mainly due to a smaller mobility value as compared to electrons: 2–4 cm$^2$/Vs for holes as compared to 30-40 cm$^2$/Vs for electrons at densities of about $10^{13}$ cm$^{-2}$).

The observation of hole transport enables using our IL-gated FETs for the determination of the ReS$_2$ band gap. In short –as the method has been discussed in detail elsewhere[19] – a change in gate voltage $\Delta V_G$ (or, more precisely, in $V_{REF}$) induces a change in Fermi energy $\Delta E_F$ and in carrier density $\Delta n$ that are related as $e\Delta V_G = \Delta E_F + e^2\Delta n/C_G$. When the Fermi level is shifted across the gap of a semiconductor, the last term can be neglected both because $\Delta n$ vanishes (or is very small: in the gap only the states originating from defects in the material are present) and the geometrical capacitance to the gate $C_G$ in IL FETs is very large[28]. As the threshold voltages for electron and hole accumulation $V_{Te}$ and $V_{Th}$ coincide with having the Fermi level respectively at the bottom of the conduction band or at the top of the valence band, their difference directly gives us the value of the band gap $e\Delta V_G = e(V_{Te} - V_{Th}) = \Delta$. To determine the gap of ReS$_2$, therefore, it suffices to measure the electron and hole threshold voltages in our IL-gated FETs and take their difference. According to its definition, the threshold voltage is obtained by looking at the $I_{SD}$-vs-$V_{REF}$ just above threshold and by finding the value of $V_{REF}$ for which $I_{SD}$ extrapolates to zero[29]. This procedure –which can be performed on a same device for different values of $V_{SD}$ to improve the accuracy with which the threshold voltages are determined[19]– is illustrated in figure 2(c). It gives $\Delta_{ReS2} = 1.41 \pm 0.05$ eV (the error is estimated by comparing the values extracted at different $V_{DS}$). Other devices give slightly larger values, but still smaller than 1.5 eV.

It is important to emphasize that several mechanisms originating from device non-idealities of different kind may affect the value extracted for the band gap[10]. For instance, a large contact resistance may strongly suppress the current when the gate voltage is above threshold, so that the threshold voltage inferred from the measurements of $I_{SD}$-vs-$V_G$ is actually larger than in the ideal case. In the presence of a very substantial density of in-gap states, $e(V_{Te} - V_{Th})$ would also be larger due to the electrostatic potential needed to fill these states (see above). All these mechanisms, however, can only lead to an overestimate of the gap (and not to an underestimate), so that –strictly speaking– the quantity $e(V_{Te} - V_{Th})$ represents an upper limit for the actual band gap. Based on these considerations, we conclude from our data that the band gap of $ReS_2$ is $\Delta_{ReS2} = 1.41 \pm 0.05$ eV or less. This value is smaller than the 1.5 eV direct optical transition observed in the PL spectra of bulk $ReS_2$[17], which has led to the claim that bulk $ReS_2$ is a direct band gap semiconductor. It matches well the ~1.4 eV indirect optical transition present in the PL spectrum recently reported in ref. [18], and in optical absorption spectra reported a couple of decades ago[30]. Our measurements therefore confirm the conclusion that $ReS_2$ has an indirect gap of approximately 1.4 eV.

Although the indirect nature of the gap in $ReS_2$ is not ideal for opto-electronic applications, the energy difference between indirect and direct gap in this material is only 100 meV. This is much smaller than in other more common semiconducting TMDs, such as $MoS_2$ and $WS_2$, in which the same difference is typically 0.5 eV or larger[11,13]. Therefore, it may be still possible to realize devices using bulk $ReS_2$ that could not be made to work with thick crystals of other semiconducting TMDs. To explore the situation we operate the same FET devices that we have used to determine the band gap of $ReS_2$ in the ambipolar injection regime[31] (i.e., by biasing in such a way that electrons and holes are injected at opposite contacts) and search for light emission.

Figure 3 shows the output curves of one of our devices driven into the ambipolar injection regime, measured for opposite polarities of source-drain voltage $V_{SD}$. The occurrence of the ambipolar injection regime –manifested by the increase in source-drain current at the end of the saturation regime (i.e., when $|V_{DS}| \gg |V_G - V_T|$)– is visible for both polarities. It is much more pronounced for negative $V_{SD}$, i.e.

when the source drain bias is used to switch from hole to electron accumulation at the injecting contact, because of the large difference in mobility between electrons and holes. Nevertheless, even for positive $V_{SD}$, a large increase in current in the ambipolar injection regime is observed by further increasing the source-drain bias, as shown in figure 4(a). For the search of EL we bias the devices in the above regime, which is preferable to the case of negative $V_{SD}$, because only a much smaller negative gate voltage is required to reach electron accumulation, making the devices significantly more stable (as mentioned earlier, at very large negative $V_G$ the occurrence of bias stress causes time dependent drifts in the operating point of the devices, making the detection of the possible occurrence of EL and the analysis of its properties more complex).

In the experiment, we fix the voltage $V_G$ applied to the IL gate at different values and ramp up the source-drain voltage $V_{DS}$ until we are sure to have reached the ambipolar injection regime. At this point we stop increasing the bias and hold it constant for the time necessary to detect light emission (approximately 60 seconds; note that the usual CCD camera mounted on the microscope in our confocal micro-PL set up is not sensitive in the infrared and therefore cannot be used to directly observe the onset of EL) and record its spectrum. We then resume sweeping $V_{DS}$ up to somewhat higher values, to ensure that the characteristics of the devices have not changed significantly during the period in which light was measured. In the three curves measured at different values of $V_G$ in figure 4(a), the interruption of the $V_{SD}$ sweeps occurred at the values indicated in the legend and is marked by the circles of the corresponding color. Despite the presence of a small "kink" in the $I_{DS}$-vs-$V_{DS}$ curves at these positions, the data show that no significant device change occurred during the time needed to measure the spectrum of the emitted light.

The normalized spectra measured following the procedure just discussed are plotted in figure 4(b). They show that irrespective of the device bias conditions, the emitted light has always an identical frequency, and originates from a radiative process at 1.5 eV, corresponding to the direct gap that has been identified from PL measurements in previous studies[17,18,32]. The blue curve represents the PL spectrum that we have measured on the same device used to study EL. The main peak in the PL spectrum matches quite well the position of the peak in the EL spectrum. PL shows additional features (also previously reported[17,18]), such as a line just above 1.6 eV (which previous work identified as a direct transition[18]) and a faint shoulder at around 1.4 eV. We find that this shoulder, due to light emission from the indirect gap[18], is more or less pronounced depending on the device, as can be seen by comparing the PL spectrum of the device used to measure EL to the spectrum of the thicker (>150 nm) $ReS_2$ crystal also shown in figure 4b (in some cases it is nearly absent, which is probably the reason why it has been missed in ref. [17], in which $ReS_2$ was claimed to be a direct band semiconductor). Neither the 1.6 eV direct transition, nor the 1.4 eV shoulder detected in PL are visible in any of the EL measurements that we have performed.

From the density of states calculated theoretically[27] and the measured electron density we estimate that Fermi level is approximately 20 meV above the bottom (the indirect minimum) of the conduction band. The electron density needed to shift the Fermi energy into the valley corresponding to the direct transition is ~8 x $10^{13}$ $cm^{-2}$. This is nearly one order of magnitude larger than the gate-induced electron density in our devices. We therefore expect that hot carriers are responsible for the observed electroluminescence. Furthermore, we note that no EL signal is observed from the direct transition at 1.6 eV, even though in PL the 1.5 and 1.6 eV transitions exhibit comparable intensity. This behavior is consistent with an electronic population described by a thermal distribution without any major deformation due to the applied electric fields. Indeed, the energy difference between the chemical potential inside the band and the direct transition at which EL is observed is ~3kT, so that a significant density of thermally excited electrons should be expected. On the contrary, the higher lying direct transition is ~7 kT away from the Fermi level, making the expected EL intensity nearly two orders of magnitude smaller than for the transition at 1.5 eV (so that the corresponding signal is eclipsed by noise). These considerations are further supported by the fact that the measured EL intensity remains virtually unchanged when varying the source drain bias by one to two Volts.

**Conclusions**

We have realized and investigated high-quality ReS$_2$ IL FETs exhibiting ambipolar transport based on bulk crystals, and used them to probe properties of this material that are of interest for possible future opto-electronic applications. In particular we have succeeded in using these devices to extract the material band gap which we found to be 1.41 eV. Besides demonstrating once more that IL-gated ambipolar transistors are very effective to determine quantitatively the band gap of semiconducting materials, our results are fully consistent with recent optical spectroscopy[18] work that reported an indirect band gap of approximately 1.4 eV for bulk ReS$_2$, smaller than the 1.5 eV direct gap. We have further shown that, despite its indirect band gap, ReS$_2$ crystals exhibit electroluminescence when IL transistors are driven in the ambipolar injection regime. Light is emitted at an energy corresponding to the direct gap of the material. In this regard, ReS$_2$ is distinctly different from more commonly studied semiconducting transition metal dichalcogenide compounds such as 2H MoS$_2$ and WS$_2$, in which electroluminescence is not observed in thick crystals, but only when the material thickness is reduced to a few monolayers. The reason for the behavior of ReS$_2$ is the small difference between the indirect and direct gap –only 100 meV –much smaller than in the other compounds just mentioned (where it typically exceeds 0.5 eV[11,13]). For this reason, it certainly appears of interest to investigate in more detail the optoelectronic properties of ReS$_2$, both in the bulk and in thin monolayers. In this regard recent work has shown that the energy of the dominant radiative transition increases as the material thickness is reduced[17,18], providing a system, having the same chemical composition and material structure, in which a broad range of energies are available for light detection and emission. However, it is still unclear whether or not there is an indirect-to-direct band gap crossover when the thickness of ReS$_2$ is reduced towards the ultimate limit[18], and on how the small energy difference between the two gaps affects the exact thickness at which the crossover takes place.

**Acknowledgments**

We gratefully acknowledge A. Ferreira for technical assistance. Financial support from the Swiss National Science Foundation (this work is part of the SNF Sinergia project "Electronic properties of new atomically thin semiconductors") and from the EU Graphene Flagship is also gratefully acknowledged.

**Author Contributions**

IGL and BAR fabricated the devices and performed the transport measurements. The photoluminescence and electroluminescence measurements were performed by NU, IGL and BAR. IGL, BAR, NU and AFM, discussed the results and analyzed the data. IGL and AFM plan the measurements and wrote the manuscript. All the authors carefully read the manuscript and commented on it.

**Figures**

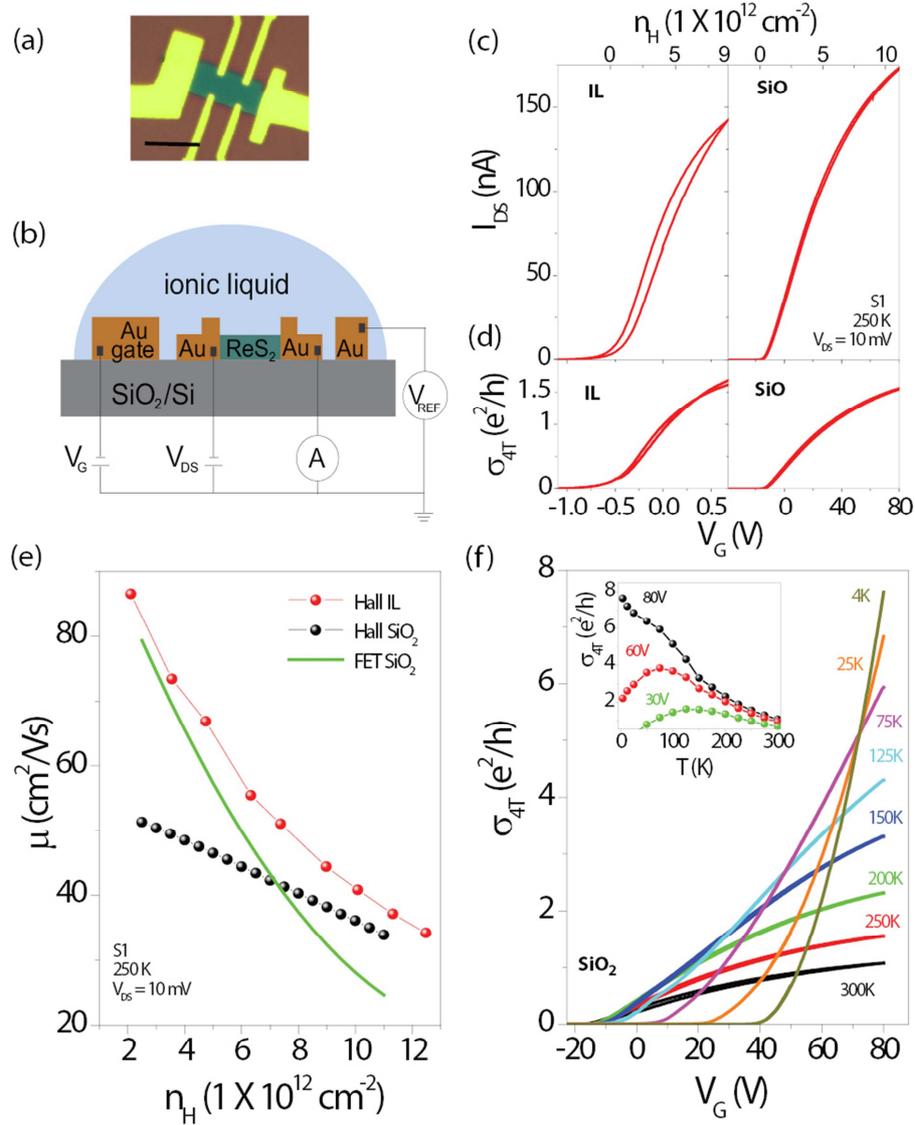

**Figure 1.** Electrical characterization of field-effect induced transport in exfoliated ReS$_2$ crystals. (a) Optical microscope image of one of our devices, with gold contacts defining a Hall bar geometry, prior to deposition of IL. The scale bar is 5 µm. (b) Schematic illustration of the device geometry, showing the electrical configuration used to operate the FETs with the IL gate (see experimental section for details). (c) and (d) Comparison of transfer curves measured using either the IL or the Si/SiO$_2$ gate at T=250K (the top panels show the drain-source current $I_{DS}$ and the bottom panels the four terminal conductivity $\sigma_{4T}$). To enable a comparison, on the top $x$-axis we report the electron density $n_h$ extracted from the measurement of the Hall resistance. (e) Comparison between the electron mobility extracted at T = 250 K from the measurements performed with either the Si back-gate or the IL gate, as a function of Hall density. (f) Temperature dependence of $\sigma_{4T}$ measured as a function of back-gate voltage $V_G$ from 300 K to 4K. The inset shows the temperature dependence of $\sigma_{4T}$ at fixed different $V_G$. All the measurements were performed at a source-drain bias $V_{DS}$ of 10 mV.

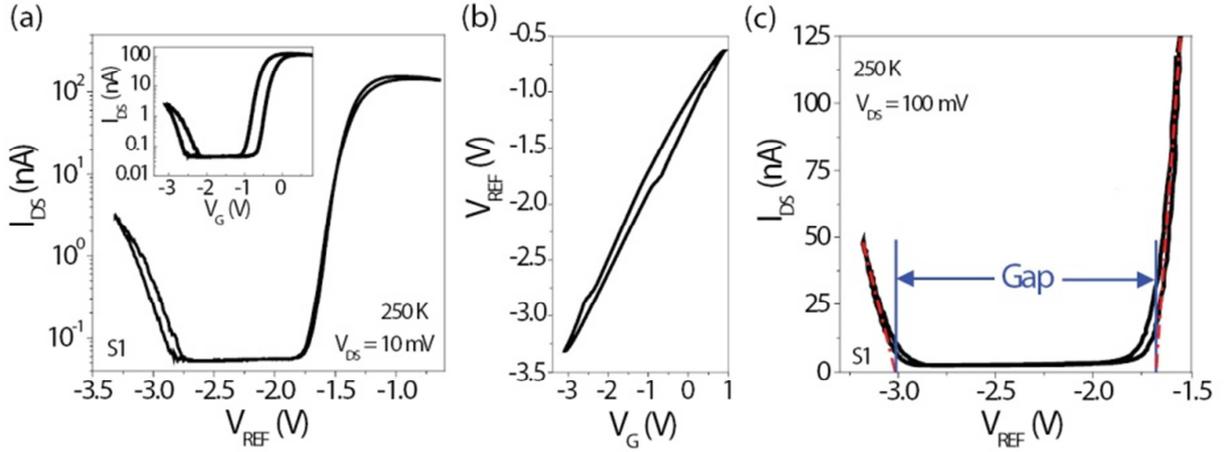

**Figure 2.** Transfer curves of an ambipolar ionic-liquid transistor realized on the surface of a thick exfoliated ReS$_2$ crystal measured at 250K. (a) Source-drain current $I_{DS}$ as a function of reference potential $V_{REF}$ (and gate voltage $V_G$, inset), measured at a source-drain voltage $V_{DS}$ of 10 mV. (b) $V_{REF}$ as a function of $V_G$. The curve shows a quasi-linear dependence of $V_{REF}$ on $V_G$, corresponding to a gate efficiency of ~65%, together with a clear hysteresis. The presence of hysteresis is consistent with the observation that the $I_{DS}$-vs-$V_{REF}$ curve is nearly hysteresis free (see panel (a)), despite the presence of a large hysteresis in the $I_{DS}$-vs-$V_G$ (see the inset of panel (a)). The observed behavior indicates that the processes responsible for the hysteresis occur at the gate-IL interface. (c) $I_{DS}$-vs$V_G$ curve in linear scale. The dash-dotted lines show the extrapolation of the $I_{DS}$-vs-$V_G$ curves to $I_{DS}$=0 A, as needed to determine the electron and hole threshold voltages. The difference between threshold voltages (multiplied by the electron charge) gives the material band gap. The data are shown for $V_{DS}$ = 100 mV; to improve the accuracy with which we determine the gap the same procedure is repeated for many different values of $V_{DS}$ (as discussed in the main text).

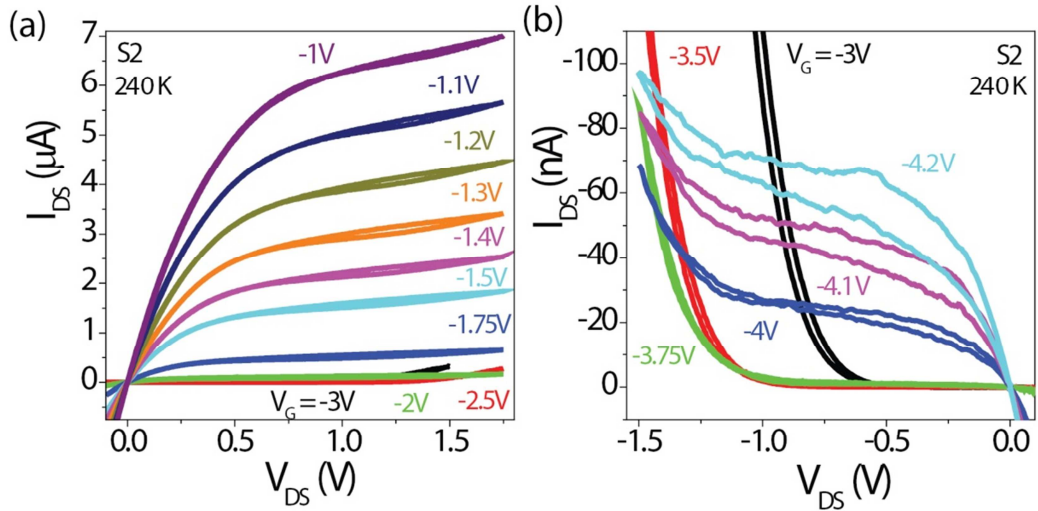

**Figure 3.** Drain-source current $I_{DS}$ vs drain-source voltage $V_{DS}$ measured on one of our ambipolar ionic-liquid gated FETs, at 240 K for different values of gate voltage $V_G$, for both positive (a) and negative (b) $V_{DS}$. In (a) the value of $V_G$ is above the threshold for electron accumulation so that at low $V_{DS}$ an electron channel is formed at the surface of the ReS$_2$ crystal. At sufficiently high positive $V_{DS}$ an upturn in $I_{DS}$ is observed at the end of the FET saturation regime, due to holes injected from the biased contact. In (b) the applied gate voltage leads to the formation of a hole accumulation channel at low bias, and for sufficiently large and negative $V_{DS}$ electrons are injected into the channel from the biased contact. The larger value of $I_{DS}$ in (a) –and the more pronounced upturn in (b) as compared to (a)– are due to the mobility of electrons that is approximately 20 times larger than that of holes.

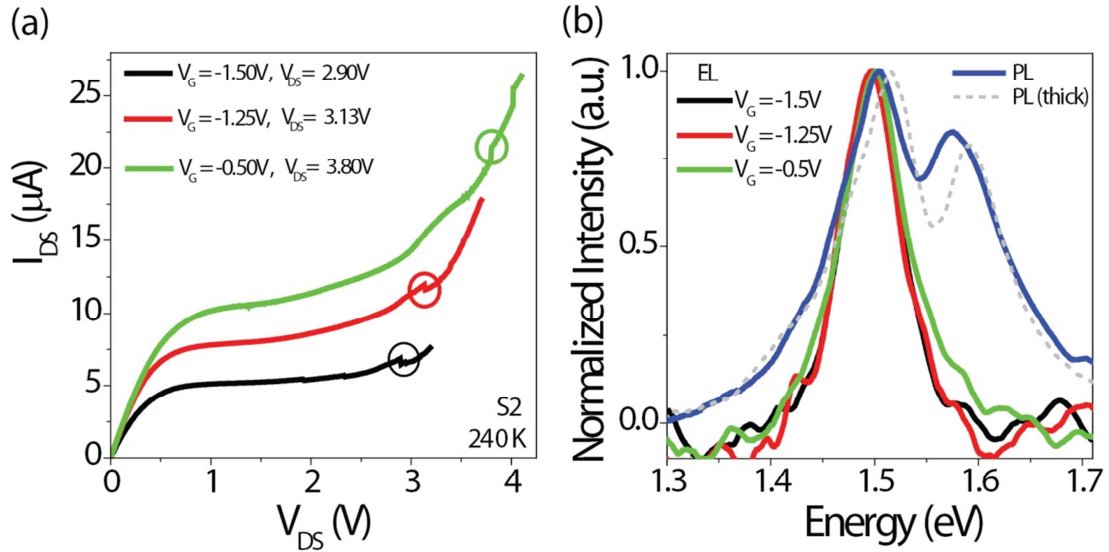

**Figure 4.** Electroluminescence (EL) from an ambipolar IL FET realized on a thick exfoliated ReS$_2$ crystal. The output curves of the device used to generate light emission due to EL are shown in (a). As compared to the data shown in Fig. 3(a), the much larger applied $V_{DS}$ leads to a much more pronounced increase in source-drain current $I_{DS}$ at the end of the saturation regime. The circles indicate the source-drain bias (value noted in the legend) at which the sweep of $V_{DS}$ is interrupted to detect the presence of emitted light and measure its spectrum. The measurements were performed at 240 K. (b) Normalized EL spectra recorded under the bias conditions shown in (a) (curves of the same colors in (a) and (b) indicate the corresponding measurements). The data show that the EL spectrum is independent of bias in the range explored. The blue line and dotted gray line are, respectively, the photoluminescence (PL) spectrum of the device in which EL was measured and that of a thick (>150 nm) ReS$_2$ crystal. The comparison between the photoluminescence and electroluminescence spectra of the device clearly shows that light emission originates from the same direct transition (the small energy difference, approximately 10 meV, found between the maxima of the PL spectra is within the spread observed in the PL spectra of different crystals; note also the faint shoulder at 1.4 eV –the lowest energy indirect transition).